\documentclass[sigconf]{acmart}

\AtBeginDocument{%
  \providecommand\BibTeX{{%
    \normalfont B\kern-0.5em{\scshape i\kern-0.25em b}\kern-0.8em\TeX}}}

 \providecommand\BibTeX{{%
  Bib\TeX}}

\copyrightyear{2024}
\acmYear{2024}
\setcopyright{rightsretained}
\acmConference[SIGIR-AP '24]{Proceedings of the 2024 Annual International ACM SIGIR Conference on Research and Development in Information Retrieval in the Asia Pacific Region}{December 9--12, 2024}{Tokyo, Japan}
\acmBooktitle{Proceedings of the 2024 Annual International ACM SIGIR Conference on Research and Development in Information Retrieval in the Asia Pacific Region (SIGIR-AP '24), December 9--12, 2024, Tokyo, Japan}\acmDOI{10.1145/3673791.3698421}
\acmISBN{979-8-4007-0724-7/24/12}

\makeatletter
\gdef\@copyrightpermission{
   \begin{minipage}{0.3\columnwidth}
     \href{https://creativecommons.org/licenses/by-nc-sa/4.0/}{\includegraphics[width=0.90\textwidth]{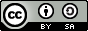}}
   \end{minipage}\hfill
   \begin{minipage}{0.7\columnwidth}
     \href{https://creativecommons.org/licenses/by-nc-sa/4.0/}{This work is licensed under a Creative Commons Attribution-ShareAlike International 4.0 License.}
   \end{minipage}
   \vspace{5pt}
}
\makeatother

\usepackage{graphicx}
\usepackage{todonotes}
\usepackage{listings} 
\usepackage{verbatim}
\usepackage{float}

\begin{document}

\title{Reproducible Hybrid Time-Travel Retrieval in Evolving Corpora}

\author{Moritz Staudinger}
\affiliation{%
  \institution{TU Wien}
  \city{Vienna}
  \country{Austria}
}
\email{moritz.staudinger@tuwien.ac.at}

\author{Florina Piroi}
\affiliation{%
  \institution{TU Wien}
  \city{Vienna}
  \country{Austria}}
\email{florina.piroi@tuwien.ac.at}

\author{Andreas Rauber}
\affiliation{%
  \institution{TU Wien}
  \city{Vienna}
  \country{Austria}}
\email{andreas.rauber@tuwien.ac.at}

\renewcommand{\shortauthors}{Moritz Staudinger, Florina Piroi, Andreas Rauber}

\begin{abstract}
There are settings in which reproducibility of ranked lists is desirable, such as when extracting a subset of an evolving document corpus for downstream research tasks or in domains such as patent retrieval or in medical systematic reviews, with high reproducibility expectations.
However, as global term statistics change when documents change or are added to a corpus, queries using typical ranked retrieval models are not even reproducible for the parts of the document corpus that have not changed. Thus, Boolean retrieval frequently remains the mechanism of choice in such settings.

We present a hybrid retrieval system combining Lucene for fast retrieval with a column-store-based retrieval system maintaining a versioned and time-stamped index. 
The latter component allows re-execution of previously posed queries resulting in the same ranked list and further allows for time-travel queries over evolving collection, as web archives, while maintaining the original ranking. 
Thus, retrieval results in evolving document collections are fully reproducible even when document collections and thus term statistics change. 
\end{abstract}



\maketitle

\section{Introduction} \label{sec_intro}

Ranked Information Retrieval has, de facto, permeated all areas of searching for information, users of nearly every search interface expecting that results are presented in some order of relevance to their query. The results returned by such an IR system to user queries are usually not reproducible when the IR system's underlying document collection changes, as is the case for any live search platform. More concretely, when documents are added to a collection or updated, global term statistics for that collection (e.g., term frequency and inverse document frequency) change. Since in ranked retrieval these global statistics contribute to the computation of relevance scores, such changes cause different result rankings. While one can argue that this lack of retrieval reproducibility is not of concern in a general (web) search setting, there are cases where, given a set of documents as a result of a search process, these search results must be traceable or reproducible at a later point in time.
This is especially true when the focus of the work is not the retrieval itself, but research based on the analysis of data that is the result of a (ranked) retrieval operation, i.e. that is, work with extracted subsets of documents from an evolving document corpus. See, for example, ``Fig 1. Corpus analysis sequence'' in \cite{10.1371/journal.pone.0184188}, where a ranked retrieval is the first step in a processing pipeline. The authors state that the same query, executed at a later time and on a document collection that has changed, has resulted in a different, larger set of documents, making the original result set not possible to recreate.

Domains that rely on reproducible and explainable retrieval results include academic search, patent retrieval, and scientific analysis of evolving collections (e.g. social media and webpages). 
An example of such a domain is systematic literature reviews (SLR), which are a type of secondary study that summarizes primary studies to answer pre-specified research questions. 
To select candidate primary studies for summarization, systematic reviews rely on constructing Boolean queries that retrieve them~\cite{hausner2015development,Higgins2019}.
The usage of Boolean queries in SLR ensures that any search can be replicated and further allows to validate the findings of such a study. For this, the documents are further filtered based on the document timestamps, to allow for consistent retrieval in evolving document collections. 

Although Boolean retrieval allows for the replication and reproduction of search results, they usually return large sets of documents, of which many are not relevant to the user's actual information need. Therefore, ranking documents according to their relevance is preferred, which is currently done with neural rerankers on the result set of a boolean query to prioritize documents for downstream tasks~\cite{wang_prioritisation2023}.

Search systems like  CORE~\cite{knoth_core_2012}, Google Scholar\footnote{\url{https://scholar.google.com/}}, or Semantic Scholar~\cite{jones_artificial-intelligence_2015}
do not provide functionality that allow researchers working with scientific content to make the results of their queries reproducible. Using APIs to interact with such systems is another form of issuing queries for retrieving data, where there is no information on how the query is further processed nor on how the results are ranked, if at all.

Reproducibility of IR ranked results, as argued for above, necessitates keeping track of all changes. This is a highly challenging task, considering the rate of change in the document collections\footnote{See ``Overall edit volume'' at \url{https://en.wikipedia.org/wiki/Wikipedia:Statistics}, or monthly submission on Arxiv\url{https://arxiv.org/stats/monthly_submissions}}.
To overcome these limitations, when working on topics that rely on the analysis of scientific publications, researchers use Boolean retrieval tools from systems like Scopus\footnote{\url{https://scopus.com/}} or PubMed\footnote{\url{https://www.ncbi.nlm.nih.gov/pubmed/}}, or more novel tools like Cruise Screening~\cite{kusa_cruise}, where tracking and sharing literature search results is possible.

In this work, we present a hybrid IR system that combines a classic, rank-based retrieval engine with a temporal, column-store-based search engine that allows for reproducing search results. 

The remainder of this paper is structured as follows. Section~\ref{sec_related}, provides a brief review on reproducibility in IR, time-travel and temporal information retrieval, while it also introduces the Research Data Alliance Dynamic Data Citation Guidelines and column-store-based retrieval.
Section~\ref{sec:architecture} describes the architecture of our proposed hybrid system. 
Section~\ref{sec:repRetrieval} shows how we reproduce retrieval results.
Section~\ref{sec:evaluation} evaluates indexing speed, query response times, score correctness, and storage overhead, followed by conclusions in Section~\ref{sec_conclusions}.
The system is released as Open Source Software\footnote{\url{https://anonymous.4open.science/r/Hybrid-Information-Retrieval}}.

\section{Related Work} \label{sec_related}
\subsection{Reproduciblity in IR}
Information Retrieval research that looks at comparing retrieval models is centered around benchmarking activities which make reproducibility a priority and a criterion for solid research. Standardised evaluation procedures that involve document corpora and tasks to be solved have become part of the IR research foundation. Retrieval evaluation campaigns like CLEF\footnote{Conference and Labs of the Evaluation Forum, \url{http://www.clef-initiative.eu/}}, TREC\footnote{The Text REtrieval Conference (TREC), \url{https://trec.nist.gov/overview.html}}, NTCIR\footnote{NII Testbeds and Community for Information access Research, \url{https://research.nii.ac.jp/ntcir/index-en.html}}, or, more recently, FIRE\footnote{Forum for Information Retrieval Evaluation, \url{http://fire.irsi.res.in}} have notably shaped this research area since the early 2000, and continue to do so. 
Efforts to standardize and automatize the retrieval evaluation setup have intensified as the number of benchmark and types of data, as well as retrieval and evaluation models are now more diverse than even a decade ago. There are now several platforms that aim to a standardization of the IR experiment evaluation and reproduction, to make IR systems comparable across domains. We mention, for example, ir\_datasets~\cite{macavaney_simplified_2021}, pybool\_ir~\cite{scells_pybool_ir_2023}, Pyserini~\cite{lin_pyserini_2021}, TIREx~\cite{frobe_information_2023}, and ir-metadata~\cite{breuer_ir_metadata_2022}.

While reproducibility of IR experiments is well catered for, with various longitudinal studies of improvements in the field \cite{yan_ictir16, arm_cikm09}, the situation changes when IR engines are used as tools for data selection, retrieving data, which is then used in further research, usually outside of the IR domain. To allow reproducible experiments in this case, researchers rely mainly on static datasets as snapshots or on Boolean retrieval \cite{pohl_extended_2010} to extract the necessary datasets for their research. 
For fields such as systematic review automation Boolean retrieval result sets are, in a second data selection step, ranked by their relevancy~\cite{karimi_csiromed_nodate}, and the resulting list is cut-off, the retained entries being further investigated in downstream tasks such as text summarization or scientific entity extraction.

\subsection{Time Information in IR}%
Using temporal information about the changes in the document collection, especially for fast changing document corpora, allows recency-based or time-dependent retrieval result ranking~\cite{kanhabua_temporal_2016}. 

Currently, for document collections where the document creation time or the harvesting time is available, reproducing search results at a later time is done by post-filtering the result set by their timestamp.
More complex solutions include index partitioning by document timestamps \cite{nandi_lifespan-based_2015} or optimizing the use of data shards and index updates to improve IR results for temporal data \cite{anand_temporal_2011, anand_index_2012}. These methods try to alleviate the effect that changes in the collection's term statistics have on later retrieval experiments, since these changes are not captured in the indexes or shards.
Our proposed system is designed to keep track of the collection term statistics changes by being augmented with a database (in our case a column-base retrieval system, see Section \ref{sec:cbs}) which allows to recreate the exact term statistics of the document collection at given point in time, reducing the overhead of shard synchronization or slicing maintenance.
With such a solution, we show that the performance of the primary IR system (the one we augment with a database) is not affected, as this system only has to export corpus and term statistics to the database.

\subsection{Columnstore-Based Retrieval}\label{sec:cbs}
While not commonly used in traditional IR systems, Chaudhuri et al. investigate the use of \textit{RDBMS} data structures for IR~\cite{Chaudhuri05} . 
With the advances in column store databases, this topic was picked up again. Gehrke~\cite{bjo_vldbws09} highlighted opportunities and challenges with Columnstore-Based Retrieval and took the first step towards a hybrid system. 

M\"uhleisen et al.~\cite{raey, Muhleisen2014} subsequently presented an IR system implemented on a column store database. They showed that sparse retrieval models such as BM25 can be transformed into SQL statements that can be efficiently executed on such systems.
It is worth noting that, while columnstore databases show viable though not competitive retrieval performance, they maintain the basic characteristics of structured data stores. 
Their proposed SQL-based retrieval approach was not intended for live information retrieval systems, but for prototyping and testing new retrieval approaches without complex indexing structures.
More recently, SQL-based retrieval was used to compare different implementations of BM25~\cite{bm25variants}.

\subsection{Dynamic Data Citation}
The challenge of precisely identifying arbitrary subsets of evolving data collections is not unique to the IR community. 
The Working Group on Data Citation of the Research Data Alliance introduced 14 recommendations in 2015~\cite{rauber_data_2015, rauber_identification_2016} to ensure precise identification of subsets of evolving datasets. %
These require tracking of data changes, storage of query data, re-execution of queries, verification of data subsets, and assignment of persistent identifiers for citations. A 2021 meta-analysis by Rauber et al.~\cite{rauber_precisely_2021} reviewed implementations across various technologies, marking significant advances in dynamic data citation. 
These implementations range from file-based systems~\cite{proll_precise_2016}, over large data collection stored in a single database~\cite{staudinger_2023_soilmoisture}, to versioning a multitude of different databases~\cite{vamdc16}.
While these solutions work well and allow to cite data across different types of data sources, it is currently not possible to cite subsets of search engine results for tasks such as systematic literature reviews, without using Boolean Information Retrieval or storing the complete result sets.

In this work, we present an approach to realize these recommendations for IR systems by combining column-store-based retrieval \cite{Muhleisen2014}, with versioning strategies for Information Retrieval Systems~\cite{bosch2017reproducible} to allow precise identification and citation of ranked result lists.
Therefore, we extend the data model by M\"uhleisen et al.~\cite{Muhleisen2014} with temporal information and adapt their query and indexing scripts, to allow for computation of reproducible ranked lists. 
Compared to M\"uhleisen et al., who precomputed all document frequency values (\textit{df}), and stored them with the dictionary terms in the columnstore, we calculate term and document statistics at the time of the query execution, taking into account that the document corpora might have changed since their first processing, and thus the document statistics are changing.
\section{System Architecture} \label{sec:architecture}
Our hybrid set-up relies on Lucene as primary search engine and MonetDB as secondary search engine to reproduce ranked lists on earlier states of the document corpus. 

Documents are preprocessed by Lucene (e.g. tokenization, stemming, stopword removal steps) for their inclusion into the Lucene Index, then synchronized with the MonetDB index.
The synchronization translates the vocabulary terms and the term statistics copied to or updated in the MonetDB index, into batch database operations. The MonetDB index updates necessarily include the timestamps of the terms' availability in the Lucene index. Lucene overwrites document representations in its index when documents get updated and reprocessed, with some terms possibly becoming unavailable. In such cases, in the MonetDB representation of the index, we mark the terms as deleted, with a deletion timestamp, and a new MonetDB entry is created for the now updated document.
\subsection{Versioned Columnstore-Based Retrieval} \label{sec_method}
To allow for the reproduction of ranked lists in evolving document collections, we adapt the model proposed by M\"uhleisen to enable Versioned Columnstore-Based Retrieval~(VCBR). 
As for sparse retrieval models, the document and term statistics change whenever a document is added, updated, or deleted, the results for a given query can also change over time. Therefore, is it necessary for VCBR to monitor every change of the document collection, and store all historical states of the corpus statistics. 

\begin{figure}[ht]
  \centering
  \includegraphics[width=0.8\columnwidth]{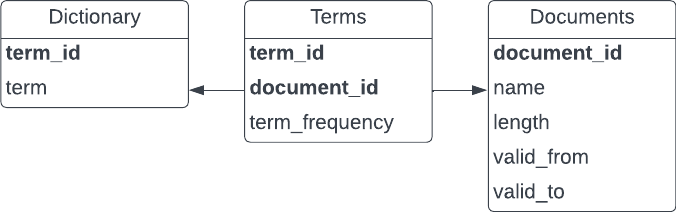}
  \caption{Excerpt of the DB schema for versioned column store based retrieval, primary keys bold}
  \label{fig:tuned_min_model}
\end{figure}

To keep all historical states, we extend the document table in the columnstore implementation of the MonetDB index with two timestamps: \textit{valid\_from} and \textit{valid\_to}. This allows us to compute the term and document statistics that were valid at a given point in time and allow document versioning (see Figure~\ref{fig:tuned_min_model}). 

Another important difference to the implementation by M\"uhleisen is, that the \emph{df} value cannot be pre-computed and stored alongside the dictionary terms, as the underlying collection evolves over time. Therefore, we calculate the \emph{df} at retrieval time for VCBR, and do not keep any pre-computed values in the database.

Our columnstore extension also includes a table for the executed queries, each with its execution timestamp, query id, a hash of the result set and the number of returned documents.
\section{Reproducing Retrieval Results}\label{sec:repRetrieval}
All queries are by default served by Lucene, returning a ranked list of $n$ documents.
Each query is stored in a dedicated table together with a persistent identified (PID), the execution timestamp, hash keys of the document IDs of the ranked result set and other descriptive metadata required for citation purposes, such as a semantic description of the sub-collection identified by the query, or the creator.
Any (subset selected via a) ranked list can subsequently be identified and cited using the PID assigned to its query.
When such an identifier needs to be resolved and the according result list be recreated, the PID is used to retrieve the query from the query store and execute it against the MonetDB index using the execution timestamp as filter to obtain the correct corpus statistics.

\lstset{language=SQL, numbers=left, tabsize=2, breaklines=true}
\begin{lstlisting}[caption= BM25 Retrieval on versioned column store with <ts> being the timestamp of original execution, label=list:model2_bm25lucene]
SELECT name, sum(bm25) as score FROM 
  (SELECT d.name, di.term, (log(1 + 
    (SELECT count(document_id) AS number FROM Documents WHERE 
      added <= <ts> AND (removed IS NULL OR removed > <ts>) - n.df + 0.5)/
      (n.df +0.5))* dt.tf / (dt.tf + 1.2 * 
      (1-0.75 + 0.75 * (d.approx_len / 
      (SELECT avg(len) AS avg_len FROM Documents WHERE 
        added <= <ts> AND (removed IS NULL OR removed > <ts>)))))) 
        AS bm25 FROM (SELECT document_id, name, valid_from, valid_to, approx_len FROM Documents 
          WHERE valid_from <= <ts> AND (valid_to > <ts> OR valid_to IS NULL)) AS d
   <...joins, grouping/ sorting...> )
\end{lstlisting}

In Listing \ref{list:model2_bm25lucene} (re-formatted and shortened for presentation purposes), we display our SQL-based BM25 variant for our versioned columnstore-based retrieval system, to recreate such a ranked list.

First, the code filters the \emph{Documents} table for those entries that were valid at the provided query time parameter <ts> (lines 4 and 10), before computing the BM25 score, which is slightly tweaked to match the computation of BM25 by Lucene (see \cite{bm25variants}). Lucene uses an approximated document length (\emph{approx\_len} in line 6, not depicted in the DB schema in Fig.~\ref{fig:tuned_min_model}), encoded using \textit{intToByte4} instead of the exact document length (\textit{len})), to improve the performance. 

After the re-execution, the hash is re-computed and verified against the hash key stored in the query store. In extremely rare cases, this can help to correct a ranked list to account for numeric discrepancies between the Lucene and MonetDB score computation.
These emerge because Lucene and MonetDB use different data types and data structures for score computation, and therefore do not produce exactly the same scores due to floating-point inaccuracies.
In rare cases (we observed 1 in 10.000) when two documents have virtually identical scores, these might be returned in swapped order in the result list. Identifying an existing discrepancy, combined with swapping document pairs with almost identical scores, allowed to correct these imprecisions in all cases in the evaluation setting.

For performing queries against any previous state of the document, it is possible to directly query the columnstore to extract a ranked list at a given point in time. Therefore, it is necessary to provide the query and a timestamp of its first execution to the column-store-based retrieval system. When no timestamp is given or available, the system queries the most recent index state.

\section{Evaluation}\label{sec:evaluation}
We evaluated performance characteristics on the first 520,000 articles of the German Wikipedia corpus dump of May 2020 (3.72GB).
Wikipedia constitutes a widely used evolving document corpus serving frequently as reference for time-travel search \cite{nandi_lifespan-based_2015}, albeit necessarily so with frozen ``dumps'' at predetermined time intervals, as any study on a live stream of updated Wikipedia documents would be difficult to reproduce with a high degree of precision.
We thus use, for the sake of this paper, a static version, simulating index update in batches, to study the performance of the hybrid system specifically with respect to efficiency of indexing and query execution as the corpus size increases.

These 520,000 documents are inserted in 26 batches with 20,000 documents each, resulting in a vocabulary of 7,787,028 unique terms and a total of 159,150,478 terms. For each batch, we measure the Lucene indexing time for Lucene and the additional overhead of transferring the information to the MonetDB index. Four pre-defined sets of 100 queries each comprising of 1, 2, 5 and 10 terms are sent to both systems to evaluate the query (re-execution) performance, correct ordering and difference in scoring, using BM25 as retrieval model, and retrieving the top-20 documents. Being aware of the complexity of evaluating the performance of IR engines and RDBMS systems~\cite{Raasveldt2018}, the numbers are predominantly meant to provide an indication of the performance of such a hybrid system.
\begin{figure}[t] 
  \centering
  \includegraphics[width=0.8\columnwidth]{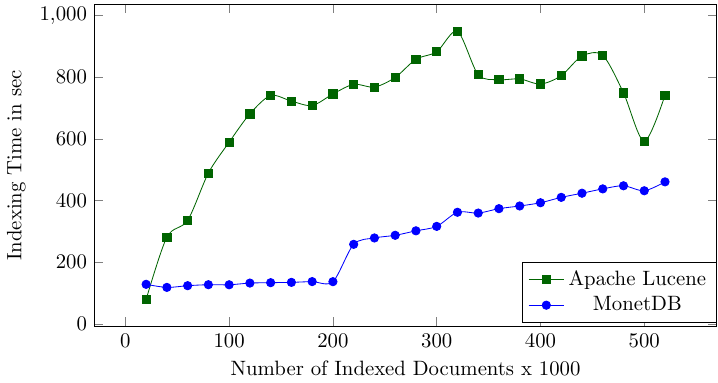}
  \caption{Indexing time Lucene (incl. parsing, stemming, stopword-filtering), and overhead for MonetDB index updates, Batch-Inserts of 20.000 documents.} \label{fig:indexingtime}
\end{figure}
\begin{figure}[t] 
  \centering
  \includegraphics[width=0.8\columnwidth]{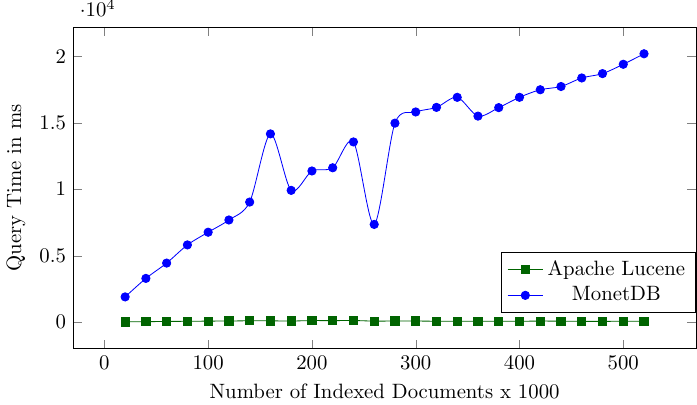}
  \caption{Query processing time over growing corpus size} \label{fig:querytime}  
\end{figure}
\begin{figure}[t] 
  \centering
  \includegraphics[width=0.8\columnwidth]{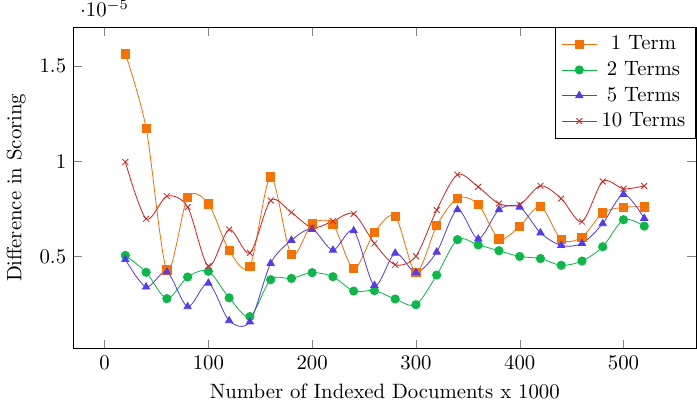}
  \caption{Evolution of the average differences between the scores returned by Lucene and VCBR due to differences in floating point operations} \label{fig:score_variation}
\end{figure}
\begin{figure}[t] 
  \centering
  \includegraphics[width=0.8\columnwidth]{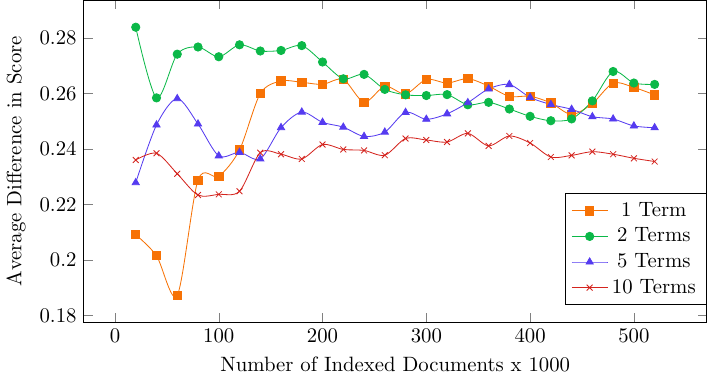}
  \caption{Average Difference of consecutive scoring results in MonetDB} \label{fig:consecutive_difference}
\end{figure}

\textbf{Indexing:} Documents are added in batches of 20.000, i.e. table updates are written only after processing and accumulating the numbers for 20,000 documents. Fig.~\ref{fig:indexingtime} depicts the resulting processing times per batch. 
Note that the times given for MonetDB represent only the additional time needed for updating the column store tables, i.e. to transfer the Lucene index information to the MonetDB based index. 
Parsing, all preprocessing such as stemming and filtering of stopwords, is performed by Lucene, resulting in seemingly higher indexing times.
The indexing time for Lucene remains rather stable as the collection size increases (approximately 13 min per 20K documents), whereas the time for updating the the column store data in MonetDB is linearly increasing, from initially 2:30 minutes to 8 minutes for the last batch.
The storage footprint of the MonetDB solution after indexing all documents, is at 4.05\,GB (Dict: 194\,MB; Docs: 39\,MB; Terms: 3.82\,GB), i.e. slightly larger than the Lucene index at 3.82\,GB. 
The hybrid system thus more or less doubles the storage requirements for the index structures, with additional storage being required for updated documents.

\textbf{Query processing times}: Evaluation of query processing times (Fig.~\ref{fig:querytime}) show a similar picture, i.e. a rather constant processing speed for the Lucene index (approx. 80\,ms per query in our set-up), compared to a linear increase over growing index size for MonetDB, ranging from 2000\,ms initially to 20 seconds against an index of 520,000 documents.
While such a performance degradation would be prohibitive in a live query processing system, it is acceptable for re-execution / reproducibility studies, as such requests are much rarer and do not need to satisfy immediate response times. Also note, that no optimization such as sharding of database tables across multiple nodes has been applied.

\textbf{Score variation:} As both systems compute the scores independently, slight differences due to floating point inaccuracies occur.
Fig.~\ref{fig:score_variation} documents the observed variations, all of which are in the range of $10^{-5}$.
As the score difference observed between two consecutive documents in the ranked list is at an order of magnitude of $10^{-2}$ (cf. Fig.~\ref{fig:consecutive_difference}), both systems return identical ranked lists.
Over the entire set of $26 \cdot 400 = 10,400$ queries, a single case occurred where two documents were sufficiently close in their score to be swapped.
This error was detected by a hash mismatch between the ranked lists. Analyzing the respective pairwise distances between subsequent documents in the list identified one sole candidate pair for swapping which resulted in a ranked list producing the correct hash value.
This error correction scheme allows to capture the rare errors possible due to floating point inaccuracies, ensuring that identical ranked lists can be used in subsequent processing steps.

\section{Conclusions} \label{sec_conclusions}
In search settings where transparency and accountability are crucial, evolving document corpora complicate the use of rank-based retrieval systems. Therefore, often simpler set-based mechanisms, such as Boolean retrieval, are employed to extract subsets of these evolving corpora in fields as scientific research or patent retrieval.

We present a hybrid system that combines both, a regular inverted index based retrieval system (Lucene) with such a versioned column store database (MonetDB).
This allows live queries to be handled by the (faster) Lucene part, whereas reproducibility of ranked lists is guaranteed by the (significantly slower) VCBR system using MonetDB. With the Columnstore implementation mimicing the behavior of Lucene, and further applying versioning to the inverted index, we enable the reproduction of the same ranked list.
This approach further allows time-travel search, by leveraging on the versioning of the VCBR engine to recreate the document statistics at a given point in time, and thus allowing querying the corpus at this time.

Future work will focus on extending the range of retrieval models that are supported and mirrored by the column-store system beyond BM25, especially to enable reproducible dense retrieval. Other research should be conducted on improving the performance of VCBR to allow to index and retrieve from larger collections. Further research should also be conducted on the reproducibility of ranked lists when using neural reranking strategies.

\bibliographystyle{ACM-Reference-Format}
\bibliography{sample-base}

\end{document}